\documentclass[fleqn,usenatbib,useAMS]{mnras}

\usepackage{newtxtext,newtxmath}
\usepackage[T1]{fontenc}
\usepackage{ae,aecompl}
\usepackage{graphicx}	
\usepackage{amsmath}	
\usepackage{multicol}        
\usepackage{bm}		
\usepackage{pdflscape}	
\usepackage{epstopdf}

\newcommand{\beq}{\begin{equation}}
\newcommand{\beqa}{\begin{eqnarray}}
\newcommand{\eeq}{\end{equation}}
\newcommand{\eeqa}{\end{eqnarray}}

\newcommand{\simgt}{\lower.5ex\hbox{$\; \buildrel > \over \sim \;$}}
\newcommand{\simlt}{\lower.5ex\hbox{$\; \buildrel < \over \sim \;$}}

\newcommand{\diff}{\mathrm{d}}
\newcommand{\sourcesizeparam}{\rho_*}
\usepackage{siunitx}

\title[Wave effect on PBH microlensing constraints]
{On the wave optics effect on primordial black hole constraints from
optical microlensing search}

\author[Sugiyama, Kurita \& Takada]
{Sunao Sugiyama$^{1,2}$\thanks{E-mail: sunao.sugiyama@ipmu.jp},
Toshiki Kurita$^{1,2}$\thanks{E-mail: toshiki.kurita@ipmu.jp},
Masahiro Takada$^{1}$
\\
$^1$ Kavli Institute for the Physics and Mathematics of the Universe (WPI),
The University of Tokyo Institutes for Advanced Study (UTIAS),
\\ The University of Tokyo, 5-1-5 Kashiwanoha, Kashiwa-shi, Chiba, 277-8583, Japan\\
$^{2}$ Department of Physics, The University of Tokyo, 7-3-1 Hongo, Bunkyo-ku, Tokyo 113-0033 Japan\\
}

\date{Accepted 2020 February 5. Received 2020 January 18; in original form 2019 June 9}

\begin{document}
\label{firstpage}
\pagerange{\pageref{firstpage}--\pageref{lastpage}}
\maketitle

\begin{abstract}
Microlensing of stars, e.g. in the Galactic bulge and Andromeda galaxy (M31), is among the most robust, powerful method to constrain  primordial black holes (PBHs) that are a viable candidate of dark matter. If PBHs are in the mass range $M_{\rm PBH}\simlt 10^{-10}M_\odot$, its Schwarzschild radius ($r_{\rm Sch}$)
becomes comparable with or shorter than optical wavelength ($\lambda)$ used in a microlensing search, and in this regime the wave optics effect on microlensing needs to be taken into account. For a lensing PBH with mass satisfying $r_{\rm Sch}\sim \lambda$, it causes a characteristic oscillatory feature in the microlensing light curve, and it will give a smoking gun evidence of PBH if detected, because any astrophysical object cannot have such a tiny Schwarzschild radius. Even in a statistical study, e.g. constraining the abundance of PBHs from a systematic search of microlensing events for a sample of many source stars, the wave effect needs to be taken into account. We examine the impact of wave effect on the PBH constraints obtained from the $r$-band (6210\AA) monitoring observation of M31 stars in Niikura et al. (2019), and find that a finite source size effect is dominant over the wave effect for PBHs in the mass range
$M_{\rm PBH}\simeq[10^{-11},10^{-10}]M_\odot$. We also discuss that, if a denser-cadence (10~sec), $g$-band monitoring observation for a sample of  white dwarfs over a year timescale is available, it would allow one to explore the wave optics effect on microlensing light curve, if it occurs, or improve the PBH constraints in $M_{\rm PBH}\simlt 10^{-11}M_\odot$ even from a null detection.
\end{abstract}




\section{Introduction}
The nature of dark matter (DM) is one of the most tantalizing problems in cosmology and physics. Unknown stable elementary particle(s) beyond the Standard Model of Particle Physics, the so-called Weakly Interacting Massive Particle(s) (WIMP), has been thought of as a viable candidate of DM, but has yet to be detected either in direct experiments, collider experiments, or indirect searches \citep[e.g.][]{Jungmanetal:96,2018arXiv181202029H,Arina:2018zcq}. Primordial black holes (PBHs) \citep{Hawking:74,Carr:75} are alternative, viable candidate of DM \citep{Carretal:16,Carr:2019yxo,Sasaki:2018dmp}.  Recently the PBH DM scenario has got attention again, partly because of recent claims that PBHs of
$10M_\odot$ mass scales can be progenitors of binary black holes whose gravitational wave have been detected by the LIGO/Virgo experiment \citep[e.g.][]{Sasakietal:16,Birdetal:16}.

Given these growing interests, there are many observational attempts to search for or constrain PBHs of various
mass scales \citep{Carretal:10,Carretal:17,Inomataetal:17,2019arXiv190410971S}. Gravitational microlensing is the most powerful, robust method of constraining PBHs \citep{Paczynski:86,Griestetal:91}, because it is a gravitational  effect and can probe mass of a lensing compact object, if detected, regardless of whether or not the lensing object is visible \citep[][for future prospects]{2019arXiv190304425B,2019arXiv190304424A}. In such a microlensing search we should keep in mind a discovery potential: if we have an even single, secure candidate of microlensing event indicating a mass scale of $M\simlt $ a few $M_\odot$ and if the counter object is confirmed as a black hole (or extremely invisible object) based on any follow-up, deep observation in various wavelengths, it can be a smoking gun evidence of PBH because any supernova explosion or other astrophysical process cannot make such a light-mass BH of $\simlt
\mbox{a few~}M_\odot$.
The pioneer work was done by the MACHO and EROS experiments that used monitoring observations of stars in the Large Magellanic Cloud to search for microlensing events and then obtained upper bounds on the abundance of compact objects over a wide range of mass scales $M\simeq [10^{-7},10]M_\odot$ \citep{Alcocketal:96,Alcocketal:00,EROS:07}. This constraint was recently updated in \citet{2019PhRvD..99h3503N} that used the public OGLE microlensing events \citep{2017Natur.548..183M}.

\citet{Niikuraetal_PBH:17}, where authors of this paper are co-authors, used the new wide-field prime-focus camera at the 8.2m Subaru telescope, Hyper Suprime-Cam (HSC), to carry out very dense cadence observation (2~min cadence) of the Andromeda galaxy (hereafter M31). Thanks to the wide field-of-view and large aperture of HSC/Subaru, they were able to monitor many stars in M31 (about 770~kpc in distance or 24.4~mag for the distance modulus) and to
search for microlensing events of much shorter timescales than previously done. They found one possible PBH microlensing event compared to the theoretically-expected number of events up to 1000 events if PBHs make up all DM in the Milky Way and M31 halo regions. The results were then translated into most stringent upper bounds on the abundance of PBHs over the range of mass scales, $M_{\rm PBH}\simeq [10^{-11},10^{-7}]M_\odot$. As discussed in \citet{Niikuraetal_PBH:17}, there is a fundamental limitation to constrain PBHs in $M\simlt 10^{-11}M_\odot$, from an optical microlensing search,  due to the finite source size effect and the wave optics effect \citep[see][for similar discussion]{2018JCAP...12..005K,2018arXiv181201427B}.

There are several earlier works discussing the wave optics effect on gravitational lensing \citep{1992grle.book.....S,Gould:92,Nakamura:98,1999PThPS.133..137N,TakahashiNakamura:03,MatsunagaYamamoto:06,2018PhRvD..97j3507N}. These considered lensing of gamma-ray burst or gravitational waves, which are in much shorter or longer wavelengths than optical light. If PBHs are lighter than $10^{-10}M_\odot$, the
Schwarzschild radius becomes comparable with or even shorter than optical wavelengths, then we cannot ignore the wave optics effect on microlensing due to interference and diffraction effects. In an extreme case even a PBH, e.g. if lighter than $10^{-11}M_\odot$, cannot bend the path of optical light from a star. The HSC/Subaru data of M31 was the first kind of data to realize that the wave optics effect can be important for optical microlensing observation.

Hence the purpose of this paper is to study the effect of wave optics on the optical microlensing search, with particular focus on the Subaru HSC M31 data in \citet{Niikuraetal_PBH:17}. For comprehensiveness, we also study the effect of finite source size on microlensing \citep{WittMao:94,CieplakGriest:13}\footnote{\citet{MatsunagaYamamoto:06} also studied both effects of wave optics and finite source size on gravitational lensing phenomena, but their study is more general, and did not discuss the consequences for optical microlensing search}. The wave effect itself for optical microlensing is quite interesting because it gives a direct evidence of PBH, if the effect is measured from the microlensing light curve even for a single event,  because any astrophysical object cannot have such a tiny Schwarzschild radius (their physical size is bigger than the light wavelength). After carefully studying the two effects on microlensing light curve, we will discuss how these affect the microlensing constraints on the abundance of PBHs in such a light mass range. Our study will give a more quantitative study of the results in \citet{Niikuraetal_PBH:17}. We will also discuss how these constraints can be improved by using a bluer optical data, such as $g$-band data, than used in \citet{Niikuraetal_PBH:17}, because the wave effect would be smaller in shorter wavelengths.

The structure of this paper is as follows. In Section~\ref{sec:basics} we review the effects of wave optics and finite source size on microlensing in optical wavelengths. In Section~\ref{sec:implication_Subaru}, we discuss the implications of wave effect and finite source size effect on the PBH constraints obtained from the Subaru HSC microlensing search of stars in M31. In Section~\ref{sec:discussion} we will give a discussion of how the optical microlensing constraints on PBHs can be improved if a microlensing search based on bluer-filter data ($g$-band filter) is used. We then give conclusion in Section~\ref{sec:conclusion}. Throughout this paper we adopt the natural units, $c=1$ ($c$ is the speed of light).

\section{The wave optics and finite source size effects on PBH microlensing}
\label{sec:basics}

In this section, we review the effects of wave optics effect and finite source size on PBH microlensing, following the papers \citep{Nakamura:98,TakahashiNakamura:03,MatsunagaYamamoto:06}

\subsection{Microlensing basics for a point source}

The characteristic angular scale of microlensing for a star in M31 due to a PBH is the Einstein radius on the sky
\citep{Paczynski:86} \citep[also see][]{1992grle.book.....S,2006glsw.conf...91K}:
\begin{align}
\theta_{\rm E}&\equiv \frac{R_{\rm E}}{d_{\rm L}}\nonumber\\
&\simeq 10^{-3}\mu{\rm as}\left(\frac{M_{\rm PBH}}{10^{-10}M_\odot}\right)^{1/2}
\left(\frac{d_{\rm S}}{d_{\rm M31}}\right)^{-1/2}\left(\frac{1-x}{x}\right)^{1/2},
\end{align}
where $R_{\rm E}$ is the Einstein radius defined as
\begin{align}
R_{\rm E}\equiv \sqrt{4G M_{\rm PBH}d_{\rm L}\left(1-\frac{d_{\rm L}}{d_{\rm S}}\right)},
\end{align}
$d_{\rm L}$ and $d_{\rm S}$ are distances to a lensing PBH and a source star, respectively, $M_{\rm PBH}$ is the PBH mass, and
$x\equiv d_{\rm L}/d_{\rm S}$. We assume that a source star is at the distance to M31 for which we assume $d_{\rm M31}=770~$kpc, and
plugged in $M_{\rm PBH}=10^{-10}M_\odot$
into the above equation as a working example.
If a source star is closer than $\theta_{\rm E}$ in separation from a lensing PBH on the sky, the source star is multiply imaged by its lensing. However, as implied by the above equation, the separation between multiple images is too small to be
resolved by an optical telescope -- this phenomena is ``microlensing''.
What is observed instead is a magnification of the total flux of two images relative to that of the original single image. The microlensing magnification
for a point source, under the geometrical optics approximation (i.e. when ignoring the wave optics effect),
is given as
\begin{align}
	A_\mathrm{nw}^\mathrm{p}(u) = \frac{u^2+2}{u\sqrt{u^2+4}},\label{eq:p-nw-magnification}
\end{align}
where $u$ is the dimension-less impact parameter between lens and source in units of the Einstein radius, the subscript
``nw'' denote {\it no wave effect}, and the superscript ``p'' denotes {\it point source}.
If $u\le 1$, a source star is multiply imaged, and the total magnification
$A\ge 1.34$ (i.e. the observed flux is brightened by a factor of 1.34 compared to the original brightness), which we will often consider a clear event of microlensing
phenomena to be detected in the following.

Because a PBH and a source star have a relative motion with respect to an observer, it causes the observed flux of a source star  to vary with observation epoch (time), leading a characteristic light curve of the observed star flux. In this way a microlensing event can be identified from the observation and is distinguishable from other variable stars, as done in many experiments \citep[e.g.][]{Alcocketal:00,EROS:07,Sumietal:03,2017Natur.548..183M}.
A typical timescale of the microlensing light curve can be estimated from a crossing time of the Einstein radius for a lensing PBH with respect to a source star:
\begin{align}
t_{\rm E}&\equiv \frac{R_{\rm E}}{v}\nonumber\\
&\simeq 9.8~{\rm min}~\left(\frac{M_{\rm PBH}}{10^{-10}M_\odot}\right)^{1/2}
\left(\frac{v}{200~{\rm km/s}}\right)^{-1}\nonumber\\
&\hspace{8em}\times\left(\frac{d_{\rm S}}{d_{\rm M31}}\right)^{1/2}
\left(x(1-x)\right)^{1/2},
\label{eq:tE}
\end{align}
where $v$ is the relative velocity for a observer-lens-source system \citep{Niikuraetal_PBH:17}. Here we assumed
$v=200~{\rm km/s}$ for a typical velocity of PBH in the halo regions of Milky Way and M31 as implied by their rotation curves (any object following the gravity in the Milky Way and M31 halo regions should have a similar velocity structure). More exactly speaking, the velocity relevant for a microlensing light curve is the velocity component on the  two-dimensional plane perpendicular to the line-of-sight direction, and one needs to take into account variations in the velocity component in the plane, as well as the dependence of lens distance. For this reason, the timescale of light curve even for a fixed mass PBH has a wide distribution \citep{Niikuraetal_PBH:17}. Nevertheless,
the above equation implies that the light curve timescale is sensitive to a mass of lensing object; a PBH of $10^{-10}M_\odot$ mass scales would give a minute timescale, while
a lensing object of solar mass scales would give a few months timescale for its light curve, e.g. as shown in MACHO experiment \citep{Alcocketal:00}. In other words, if we can identify a {\it secure} microlensing event that has a very short timescale such as minute timescale, it would be a smoking gun detection of light-mass PBH such as $10^{-10}M_\odot$, because any astrophysical processes cannot produce a compact object of such small mass scales.

\subsection{Effect of wave optics}
\label{wave-effect}

If the Schwarzschild radius of a lensing object is sufficiently larger than the wavelength of light used in a microlensing observation,
the microlensing event is well described by ``geometrical optics approximation''
\citep{1992grle.book.....S}.
On the other hand,
if the Schwarzschild radius becomes comparable with or shorter than the light wavelength,
we need to take into account the wave optics effect (interference and diffraction effects) on the lensing magnification.
In the following, we briefly review the wave optics effect on microlensing for a point source, following \citet{TakahashiNakamura:03}.

The wave effect is characterized by the parameter ``$w$'', defined as
\begin{align}
	w &\equiv \frac{8\pi GM_{\rm PBH}}{\lambda}\nonumber\\
	&=
	4 \pi \frac{r_\mathrm{Sch}}{\lambda} = 5.98 \left(\frac{M_{\rm PBH}}{10^{-10}M_\odot}\right) \left(\frac{\lambda}{6210 \text{\AA}}\right)^{-1}\ ,
	\label{eq:w_def}
\end{align}
where $r_\mathrm{Sch}$ is the Schwarzschild radius of lensing PBH,
and $\lambda$ is the characteristic wavelength of light in an observation (here we assumed 6210\AA corresponding to a central wavelength of $r$-band in the Subaru telescope as our default choice).
Once we fix the light wavelength, the wave parameter depends only on the mass of PBH,
so the wave effect is independent of the distance of PBH.
This is because wave effect is not a geometrical effect which depends on the path of light, but rather a local effect around the lensing object.

The magnification including the wave optics effect,
$A^\mathrm{p}_\mathrm{w}(w,u)$, is given in \citet{1992grle.book.....S} \citep[also see][]{Nakamura:98,1999PThPS.133..137N,TakahashiNakamura:03} as
\begin{align}
	&A^\mathrm{p}_\mathrm{w}(w,u) = \frac{\pi w}{1-e^{-\pi w}}\left|{}_1F_1\left(\frac{i}{2}w,1;\frac{i}{2}wu^2\right)\right|^2\ ,
	\label{eq:A_p}
\end{align}
where ${}_1F_{1}$ is the confluent hypergeometric function. The maximum magnification is realized by setting the impact parameter $u=0$:
\begin{align}
	&\left. A^\mathrm{p}_\mathrm{w}\right|_{\rm max} = A^\mathrm{p}_\mathrm{w}(w,0) = \frac{\pi w}{1-e^{-\pi w}},
	 \label{eq:max}
\end{align}
where the subscript ``w'' denotes ``wave effect''. From this equation we can find  that, even if microlensing occurs and if $w\ll 1$, the maximum magnification can be significantly reduced
as
$\left.A^\mathrm{p}_\mathrm{w}\right|_{\rm max}\approx 1+\pi w/2$ ($A^\mathrm{p}_\mathrm{w}=1$ means no lensing magnification) compared to $A\rightarrow \infty$ in the geometrical optics approximation case. This is because the gravitational potential induced by the lightest PBH is too weak to bent the path of optical light.

When the wavelength becomes very short compared to the Schwarzschild radius of lensing object, i.e. $w\gg 1$,
the magnification under the geometrical optics approximation is realized:
$A^{\rm p}_\mathrm{w}(w,u) \rightarrow A^{\rm p}_\mathrm{geo}(w,u)$:
\begin{align}
	A^\mathrm{p}_\mathrm{geo}(w,u) &= \frac{u^2+2}{u\sqrt{u^2+4}} \nonumber\\
    &\hspace{-1em}+ \frac{2}{u\sqrt{u^2+4}}\sin\left[w\left\{
    \frac{1}{2}u\sqrt{u^2+4}+\log\left|\frac{\sqrt{u^2+4}+u}{\sqrt{u^2+4}-u}\right|\right\}\right].
    \label{eq:A_s}
\end{align}
The second term of the above equation is a rapidly oscillatory function of $u$ for a fixed $w$ ($w\gg 1)$.
In practice the second term averages out, i.e. is irrelevant in an actual observation,
if we consider a finite source size (see below) or
if we consider a finite range of light wavelengths in an optical filter and/or a finite exposure time of an  observation.
For this reason, the second term is often ignored under the geometrical optics approximation, which becomes equivalent to Eq.~(\ref{eq:p-nw-magnification}).

\begin{figure}
	\includegraphics[width=1.0\columnwidth]{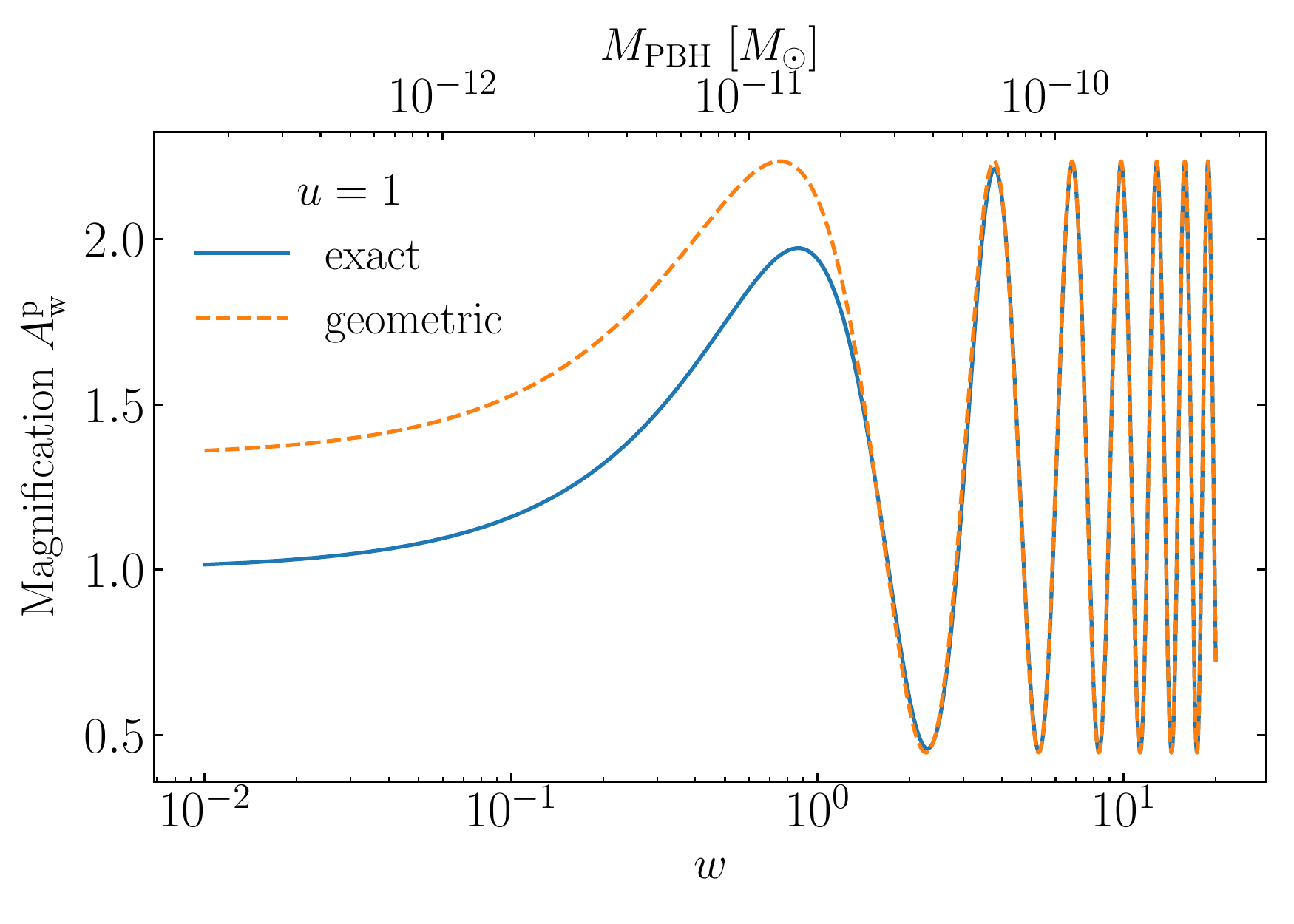}
    \caption{Lensing magnification of a point source, $A^{\rm p}$,  as a function of the ``wave parameter'', $w\equiv 8\pi M_{\rm PBH} G/\lambda
    =4\pi r_{\rm Sch}(M_{\rm PBH})/\lambda$, where $M_{\rm PBH}$ is a PBH mass, $r_{\rm Sch}$ is its Schwarzschild radius, and
    $\lambda$ is the wavelength of light used in a microlensing observation.
    For an optical-wavelength observation, the $w$ parameter in the $x$-axis can be read as PBH mass scale; the values in the upper $x$-axis correspond to PBH masses for the $r$-band filter wavelength ($\lambda=6000$\AA).
    For this plot, we fixed $u=1$ for the impact parameter in units of the microlensing Einstein radius, $u=b/R_{\rm E}=1$; it leads to $A=1.34$ for the geometrical optics limit denoted by the dashed curve. Although the curve has a rapid oscillation at $w\gg 1$, what is actually observed is the averaged magnification, e.g. over a finite exposure time, which indeed gives $A=1.34$.
 The solid curve shows the result taking into account the wave effect.   When $w\simlt 1$, where the wavelength becomes longer than the PBH Schwarzschild radius, the dashed and solid curves start to deviate from each other, meaning that
 the wave effect becomes significant. It has an asymptotic limit, $A\rightarrow 1$ at $w\ll 1$, i.e. no lensing magnification due to the wave effect.}
    \label{fig:wA}
\end{figure}

Fig.~\ref{fig:wA} compares lensing magnification taking into account wave optics with that of geometrical optics
approximation for a point source in M31. When $wu \simgt \mathcal{O}(1)$, the two curves agree with each other, which implies that
the geometrical optics approximation holds valid.
On the other hand, when $wu<\mathcal{O}(1)$, the wave effect becomes significant, leading to less magnification compared to the that of geometrical optics. This means that in the regime of $wu<\mathcal{O}(1)$ the geometrical approximation is not valid. For $wu\ll 1$ as an extreme case,
$A^{\rm p}_{\rm w}\rightarrow 1$, meaning no magnification due to the wave optics
effect.

\begin{figure}
	\includegraphics[width=1.0\columnwidth]{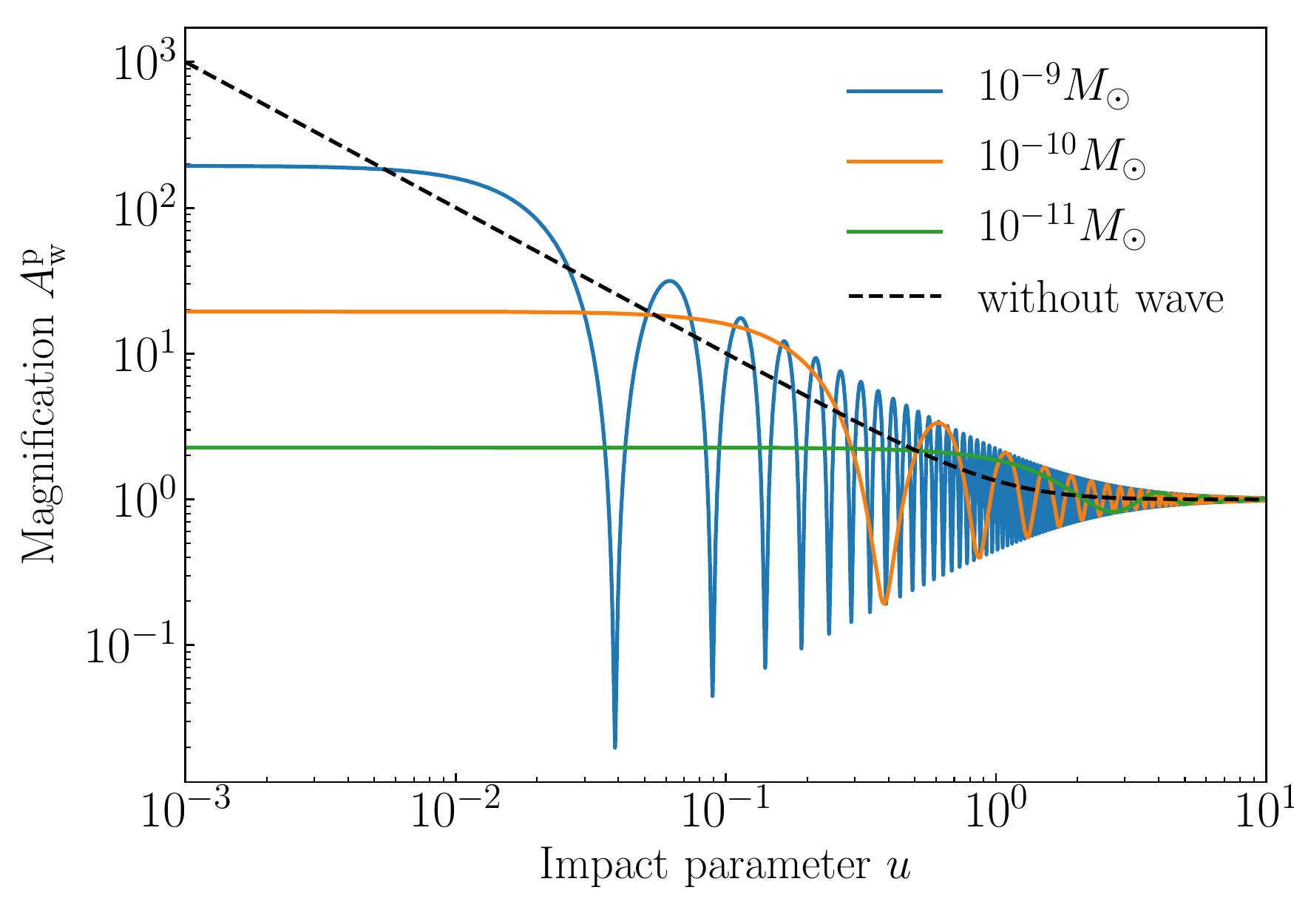}
    \caption{Lensing magnification of a point source
    as a function of the impact parameter $u$ for a lensing PBH of different mass scales.
    As in Fig.~\ref{fig:wA}, we adopted $\lambda=6000$\AA~ for the light wavelength. Each solid curve is equivalent to a light curve of microlensing, but in an actual observation we need to take into account the average of magnification over a finite exposure time and/or a finite range of wavelengths in a filter, which can be compared to the geometrical optics limit (dashed curve). Each curve becomes flattened at small impact parameters due to the wave effect. The maximum magnification at small impact parameters is lower for lighter-mass PBHs due to more significant wave effect.
    }
    \label{fig:lcp}
\end{figure}

Fig.~\ref{fig:lcp} shows lensing magnification as a function of impact parameter $u$ for PBHs of different mass scales, assuming the optical wavelength $\lambda=6000~$\AA~ as in Fig.~\ref{fig:wA}. For a microlensing system we are interested in, the impact parameter is time-varying as given by $u=u_{\rm min}+ v t$,
where $v$ is the relative velocity of lens-source-observer system ($v$ is considered constant for a microlensing phenomenon of interest), $u_{\rm min}$ is the minimum separation between lens and source,
and $t$ is time from the minimum separation. Therefore the curves in this figure are equivalent to a light curve of microlensing as a function of observation time.
As expected from Fig.~\ref{fig:wA}, the heavier the lens object is, more rapidly the lensing magnification oscillates with $u$.
This feature plays an important role when considering the effect of finite source size.
We also note the maximum lensing magnification has an asymptotic behavior at $u\rightarrow 0$: $A^\mathrm{p}_\mathrm{w}\rightarrow
{\rm constant}$, while the geometrical optics approximation (Eq.~\ref{eq:p-nw-magnification}) gives $A^\mathrm{p}_{\mathrm{nw}}\rightarrow u^{-1}$ at $u\rightarrow 0$ as explicitly shown by the dashed curve.

As obvious from Figs.~\ref{fig:wA} and \ref{fig:lcp}, characteristic features of the wave optics effect on PBH microlensing are oscillatory features in the light curve. If we can find any microlensing event, even a single event, that shows such an oscillatory feature in its light curve from an optical observation, it would be a smoking gun evidence of PBH having light mass in the range $M_{\rm PBH}\simeq [10^{-11},10^{-10}]M_\odot$, because any astrophysical, compact object cannot have such a tiny Schwarzschild radius comparable with optical light wavelength (the physical size of any compact object should be much greater than the wavelength).
However, to detect such an oscillatory feature in the light curve, we need a sufficiently dense cadence observation to well sample the light curve. This requires a careful design of the observation strategy.
Here the lower cut $10^{-11}M_\odot$ comes from the fact that there is no lensing effect for PBHs with masses below the mass limit, because of too strong wave effect.
Hence the wave effect in the optical microlensing is worth to explore from an actual observation.

We have so far assumed a static spectrum of light, and implicitly assumed that light of different paths due to lensing has an interference when taking into account the wave effect. Exactly speaking, a source might have a temporal variation in its flux, and this might prevent such an interference. Here we comment on the temporal coherence in the lensing phenomena.

Since light from a star is a black-body radiation that is characterized by its surface temperature ($T_{\rm S}$), we can naively consider that a star is a static object, i.e. has no temporal variation in the flux. Nevertheless we consider a possible shortest temporal variation or a possible shift in its light frequency; such a shift might be from 
light in an absorption line of stellar spectrum, which is caused by a motion of gas in the stellar atmosphere. Compared to this, a continuum spectrum of a star can be safely considered static (i.e. much longer time variation). The shortest timescale of such an absorption line can be estimated by the Doppler effect due to a thermal motion of gas elements in the stellar surface \citep{guenther_modern_2015}, which causes a shift in the frequency of light by 
$\Delta \nu/\nu\simeq v_{\rm th}$:
\begin{align}
\tau_\mathrm{c} &\sim \frac{1}{\Delta \nu}\nonumber\\
& \sim \frac{\lambda}{v_{\rm th}}
\simeq 
5\times 10^{-11}~{\rm sec}
    \left(\frac{m}{m_\mathrm{p}}\right)^{1/2}
    \left(\frac{T_\mathrm{S}}{6000\mathrm{K}}\right)^{-1/2}
    \left(\frac{\lambda}{6210 \text{\AA}}\right),
    \label{eq:t_doppler}
\end{align}
where $v_{\rm th}$ is the velocity of gas elements in thermal equilibrium for which we assume $m v_{\rm th}^2/2=3T_{\rm S}/2$, 
we considered the hydrogen mass ($m_{\rm p}$) for gas element for simplicity, 
we assumed $T_{\rm S}=6000~{\rm K}$ for the surface temperature as in the Sun, and $\lambda$ is a typical wavelength in the $r$-band. On the other hand, for microlensing, we want to consider an interference or superposition of light rays that arrive at an observe after traveling along the two light paths that are bent by a lensing PBH. The relevant timescale is a ``time delay'' that refers to as a difference in their arrival time of the two light paths, given in \citet{1992grle.book.....S} as
\begin{align}
    \Delta t_{\rm lens} &\simeq
    2.0\times10^{-15}\mathrm{sec}\left(\frac{M}{10^{-10}M_\odot}\right)K(u), 
\end{align}
where 
\begin{align}
    K(u) \equiv \frac{1}{2}u\sqrt{u^2+4} + \log\left|\frac{\sqrt{u^2+4}+u}{\sqrt{u^2+4}-u}\right|~ .\label{eq:time-delay}
\end{align}
Note $K(u)\sim {\cal O}(1)$ for $u<1$ in the multiple image regime. 
From Eqs.~(\ref{eq:t_doppler}) and (\ref{eq:time-delay}), we can find 
$\Delta t_{\rm lens}\ll \tau_{\rm c}$, meaning that the lensing time delay is much shorter than a possible temporal variation of light from a source star even if we consider a light around an absorption line. Thus we can safely consider a temporal coherence of a star light, and in other words we can safely consider an interference of light in the microlensing case (the use of Eq.~(\ref{eq:A_p}) is valid). We again note that the Subaru $r$-band photometry mainly measures a portion of a continuum spectrum in a black-body radiation of a star, which has much longer temporal variations, even if it has, than the lensing time delay. Thus we can safely consider that 
Eq.~(\ref{eq:A_p})
is valid for our study.

\subsection{Effect of finite source size}
\label{finite-effect}

The effect of finite source size on microlensing
is characterized by the ratio of source size to the Einstein radius on the sky \citep{WittMao:94}. We use the parameter $\sourcesizeparam$
to denote the ratio, defined as
\begin{align}
	\sourcesizeparam &\equiv   \frac{\theta_{\rm S}}{\theta_{\rm E}}=
	\frac{R_\mathrm{S}/d_\mathrm{S}}{R_\mathrm{E}/d_\mathrm{L}} \nonumber\\
	&\simeq 5.9 \left(\frac{R_\mathrm{S}}{R_\odot}\right) \left(\frac{d_\mathrm{S}}{d_\mathrm{M31}}\right)^{-1/2} \left(\frac{M_{\rm PBH}}{10^{-10}M_\odot}\right)^{-1/2} \left(\frac{x}{1-x}\right)^{1/2},
    \label{eq:U_def}
\end{align}
where $R_{\rm S}$ is the finite source size for which we assumed the solar radius.
If $\sourcesizeparam\simgt 1$, the source size effect becomes significant for the microlensing magnification. Even if $\sourcesizeparam\sim O(0.1)$,
the effect is not negligible as we will show below.
Throughout this paper we mainly consider the solar radius, $R_\odot$, for the source star size following \citet{Niikuraetal_PBH:17} that use a monitoring observation of (mainly main-sequence) stars in M31 with Subaru/HSC. Even if the physical source size is fixed, the
the source size effect varies with the relative position of lensing PBH to an observer, as given by
 $\sourcesizeparam \propto \left[x/(1-x)\right]^{1/2}$;
for a smaller $x$ (a PBH closer to an observer), $\sourcesizeparam$ becomes smaller.
This means that the source size effect is stronger when a PBH is lighter or if a PBH is nearer to the source.

To model the finite source size effect, we assume that a source star is a circular disk with constant
surface brightness. This is a simplified assumption, but is sufficient for our purpose.
Assuming that the source center is separated from a lensing PBH by $u$ (i.e. the impact parameter between the source center and lens),
we can compute the lensing magnification for such a disk-shaped star from an average of the lensing magnification (Eq.~\ref{eq:p-nw-magnification})
over the source disk:
\begin{align}
A^{\rm f}(u, \sourcesizeparam)\equiv \frac{1}{\pi \sourcesizeparam^2}\int_{|{\bf y}|\le \sourcesizeparam}\!\mathrm{d}^2{\bf y} A^{\rm p}_{\rm nw}(|{\bf u}-{\bf y}|),
\end{align}
where the integral variable ${\bf y}$ moves within the circular source disk of radius $\sourcesizeparam$, we set the origin (${\bf y}\equiv {\bf 0}$) to the source center, the superscript ``f'' in $A^{\rm f}$ denotes the ``finite source size effect'',
and
we set a lensing PBH to be located at ${\bf u}=(u,0)$ due to symmetry of the circular disk star.
Note $\int_{|{\bf y}|\in \sourcesizeparam}\mathrm{d}^2{\bf y}=\pi \sourcesizeparam^2$. When further taking into account the wave optics effect, we
replace $A^{\rm p}_{\rm nw}$ in the above equation with $A^{\rm p}_{\rm w}$ (Eq.~\ref{eq:A_p}) and then perform the numerical
integration to obtain the magnification for a given set of parameters ($u,\sourcesizeparam,w$).

\begin{figure}
	\includegraphics[width=1.0\columnwidth]{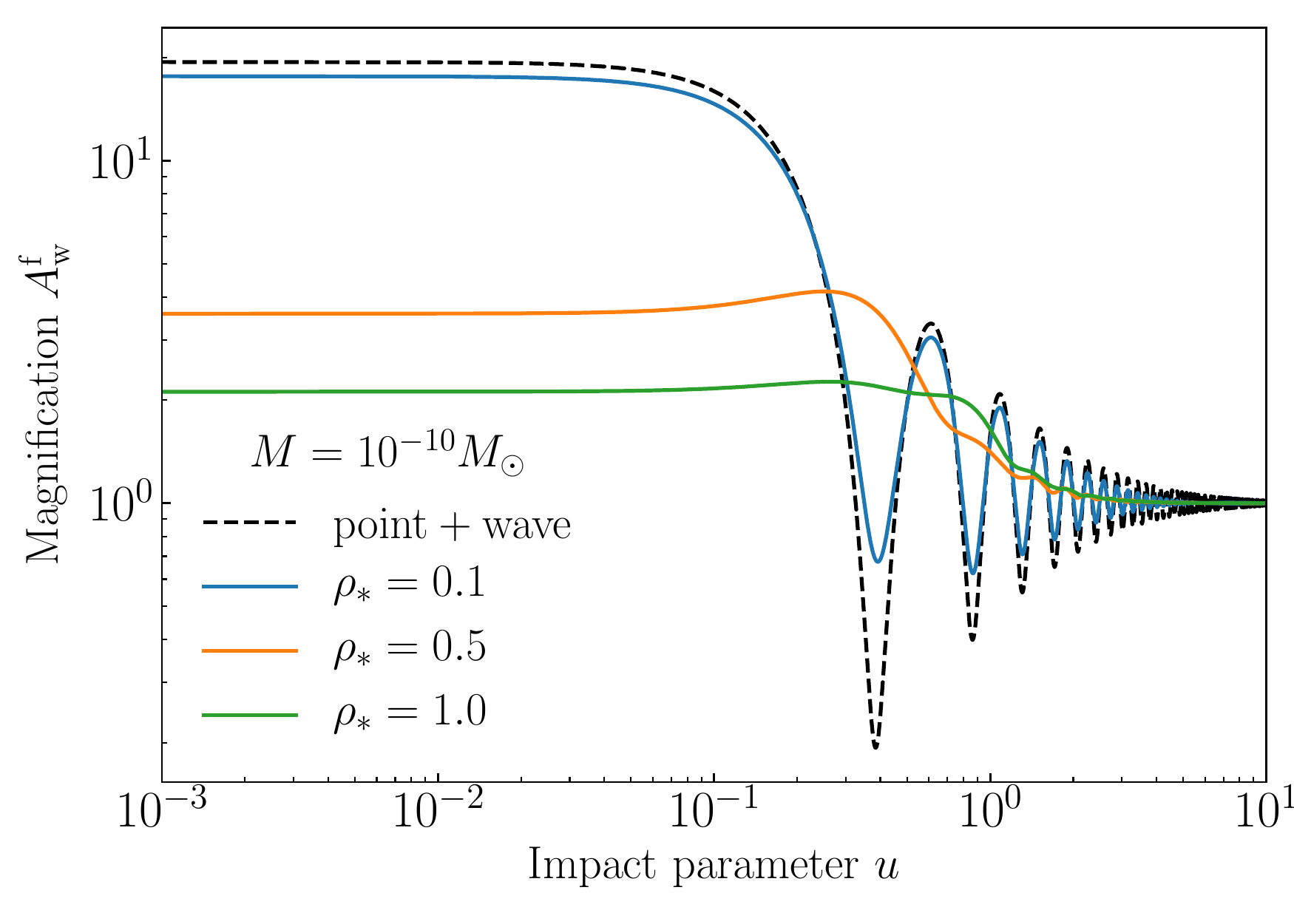}
    \caption{As in Fig.~\ref{fig:lcp}, but the solid curves show the lensing magnification when further taking into account the finite source size effect in addition to the wave effect.
    Note that the plotting range of $y$-axis is different in this plot from that in Fig.~\ref{fig:lcp}.
    We consider several cases for the source size as given in the legend.
    The source size is defined in units of the Einstein radius of a lensing PBH: e.g., $\sourcesizeparam=1$ corresponds to $\theta_{\rm S}=\theta_{\rm E}$ on the sky.}
    \label{fig:lcf}
\end{figure}

Fig.~\ref{fig:lcf} shows lensing magnification as a function of the impact parameter for different input values of $\sourcesizeparam$ corresponding to
different physical source sizes for a fixed PBH mass, $M_{\rm PBH}=10^{-10}M_\odot$.
When taking into account the finite source size, the lensing magnification becomes flattened at separations, $u\simlt \sourcesizeparam$.
For $\sourcesizeparam=0.1$, the finite source size effect is not significant, but not negligible.
On the other hand, the finite source size effect smears out oscillatory features in the light curve, because
different points in a source (disk-like shape source) have different phases in the wave optics effects of lensing magnification, and the oscillatory features are averaged out when integrating the lensing magnification over the source region.
Thus the finite source size makes it difficult to extract an oscillatory feature in the microlensing light curve. Nevertheless we should note that the finite source size effect depends on a lens distance; the effect
becomes relatively  less important for a lensing PBH closer to an observer, while the wave effect is independent of lens distance. This difference might help distinguish the wave effect from an observed microlensing light curve.

\subsection{Effect of exposure time average}
\label{sec:exposure}

For an actual observation of the PBH microlensing search, we need to further take into account the effect of finite exposure time. Here an ``exposure'' time means a time duration during which photons from a source star are collected and then the total number of photons (more exactly electrons converted from the photons) in one exposure is stored into the hard drive after closing a shutter of the camera. In the following we denote the exposure time as $t_{\rm exp}$.

We throughout this paper adopt
$t_{\rm exp}=90~\mathrm{sec}$ as our fiducial parameter following the observation in \citet{Niikuraetal_PBH:17}.
The impact parameter of a given microlensing event would be changed during the exposure time by an amount of
\begin{align}
	\Delta u &\simeq  \frac{v t_{\rm exp}}{R_E}\nonumber\\
    &= 0.15 \left( \frac{v}{200~{\rm km/s}} \right)
     \left( \frac{t_{\rm exp}}{90~{\rm sec}} \right)\nonumber\\
    &\hspace{2em}\times\left(\frac{M_{\rm PBH}}{10^{-10} M_\odot}\right)^{-1/2} \left(\frac{d_\mathrm{S}}{d_\mathrm{M31}}\right)^{-1/2}
    \left(x(1-x)\right)^{-1/2}\ ,
	\label{eq:Delta_u}
\end{align}
Thus what we can observe is the microlensing magnification averaged over the exposure time or equivalently the interval $\Delta u$ around the fiducial point of $u$.
Because of $\Delta u\propto v/R_{\rm E}$ for a fixed exposure time, the averaging effect is more significant when
$v$ is greater or $R_{\rm E}$ is smaller.
Hence the lensing magnification including the effect of
a finite exposure time can be estimated as
\begin{align}
	\bar{A}(u) = \frac{1}{\Delta u} \int_{u-\Delta u/2}^{u+\Delta u/2}\!\mathrm{d}u'~ A(u') .
	\label{eq:A_texp}
\end{align}
More exactly speaking, however, the averaged magnification needs to be computed from a given trajectory around the fiducial point $u$. In the above equation, we simply assume that the trajectory is along the path towards the center of source star (i.e. the path with the minimum impact parameter $u_{\rm min}=0$). This is a simplified treatment, but is enough for our purpose. We will use this equation to estimate the impact of finite exposure time on our results.

\begin{figure*}
	\includegraphics[width=0.46\textwidth]{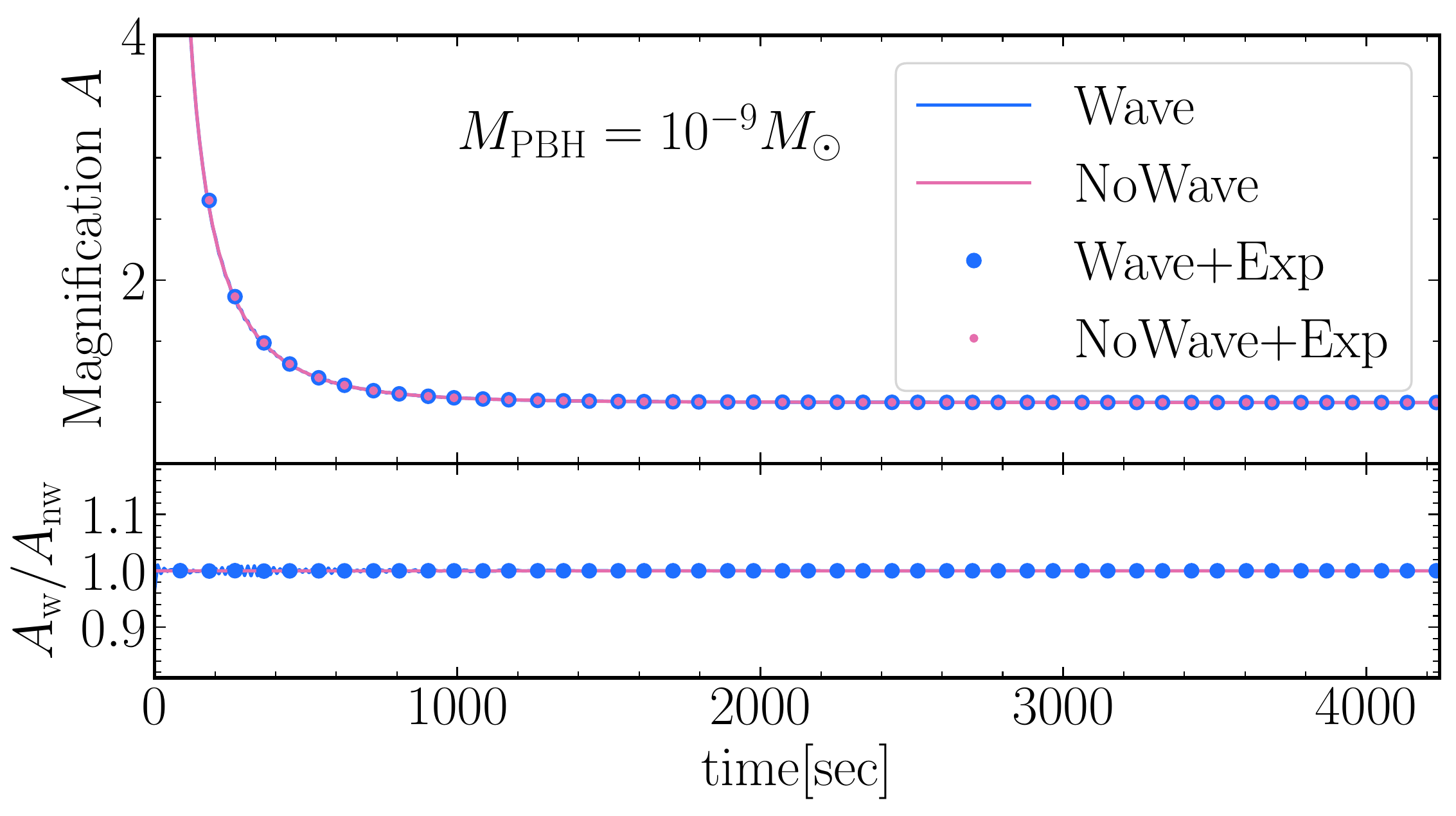}
	\includegraphics[width=0.46\textwidth]{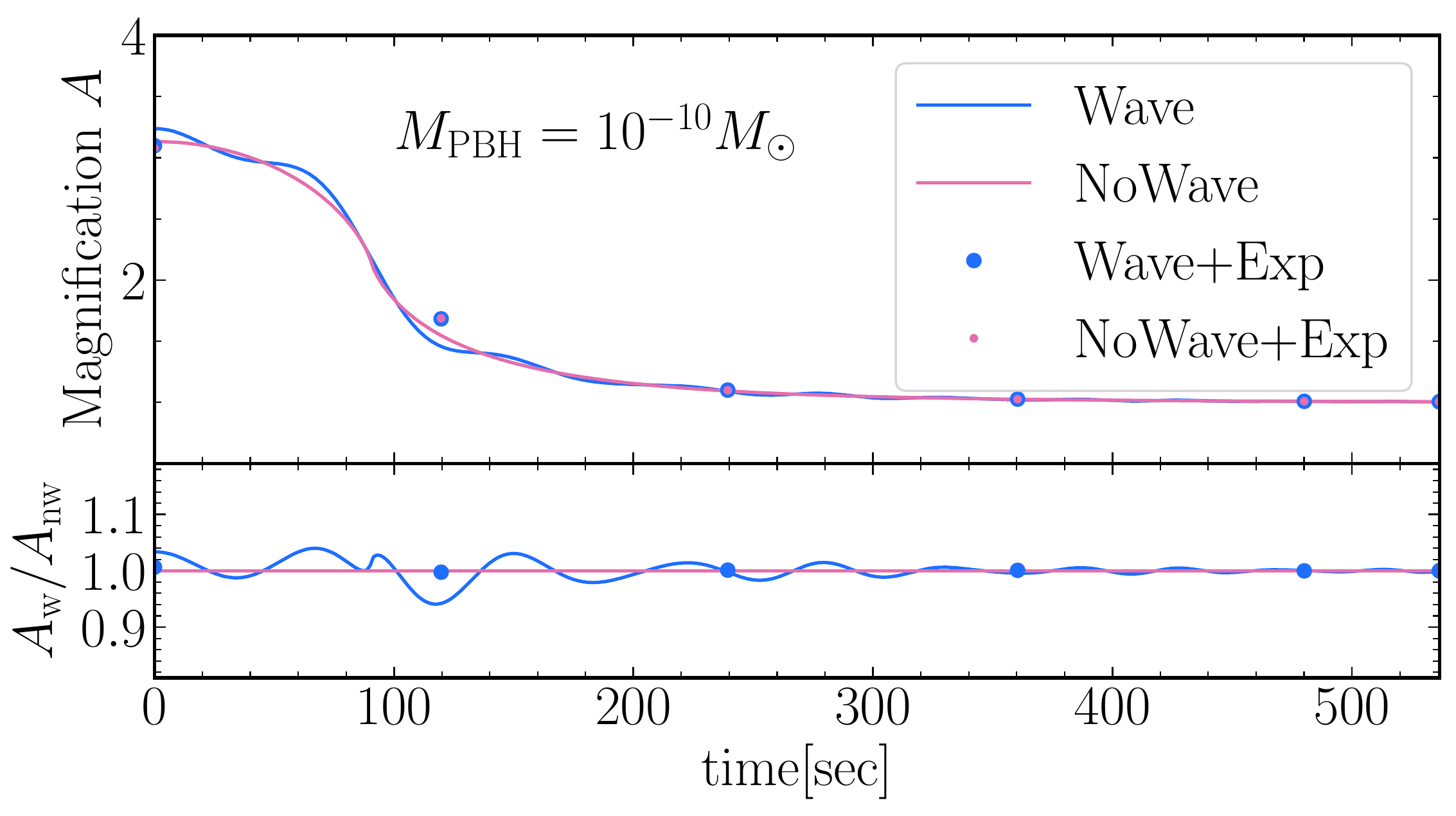}
	\caption{A simulated light curve for a microlensing due to a PBH with $M_{\rm PBH}=10^{-9}M_\odot$ (left panel) and $M_{\rm PBH}=10^{-10}M_\odot$ (right panel), for a star in M31. Here we consider a case that a lensing PBH is at distance $d_{\rm L}=10~{\rm kpc}$, has a perpendicular velocity of 100~km/s, and assume 
	an observed wavelength of $\lambda=6210$\AA as in the Subaru $r$-band and 
	a solar radius for a source star size, $R_{\rm S}=R_\odot$, as a working example. In this case the wave effect parameter $w=59$ ($5.9$) and the finite source size relative to the Einstein radius $\rho_\ast=0.21$ ($0.66$) for the case of $M_{\rm PBH}=10^{-9}M_\odot$ ($10^{-10}M_\odot$). Furthermore, we consider the discrete sampling effect in a measurement of the light curve; we here assume a 2~min cadence meaning that the light curve is sampled every 2~min as denoted by circle points. We here compare the results with and without the wave effect. The lower panels show the ratio.
	An oscillatory feature in the light curve due to the wave effect is difficult to be captured because of the finite source size effect for $M_{\rm PBH}=10^{-9}M_\odot$ and exposure time average effect for $10^{-10}M_\odot$, respectively.
}
	\label{fig:simulated_lightcurves}
\end{figure*}

Now we come back to a question of whether the wave effect on the microlensing light curve can be measured by a dense cadence of an observation such as that done in \citet{Niikuraetal_PBH:17}, where an imaging data  of M31 was taken every 2~min (90~sec for exposure plus about 30~sec readout of the data). Fig.~\ref{fig:simulated_lightcurves} shows a simulated light curve for a typical case of microlensing for such an observation. 
As can be found, an oscillatory feature due to the wave effect 
is difficult to be captured because of the insufficient cadence.

\section{Implications on the PBH constraints from the Subaru HSC microlensing search}
\label{sec:implication_Subaru}

\begin{figure}
	\includegraphics[width=1.0\columnwidth]{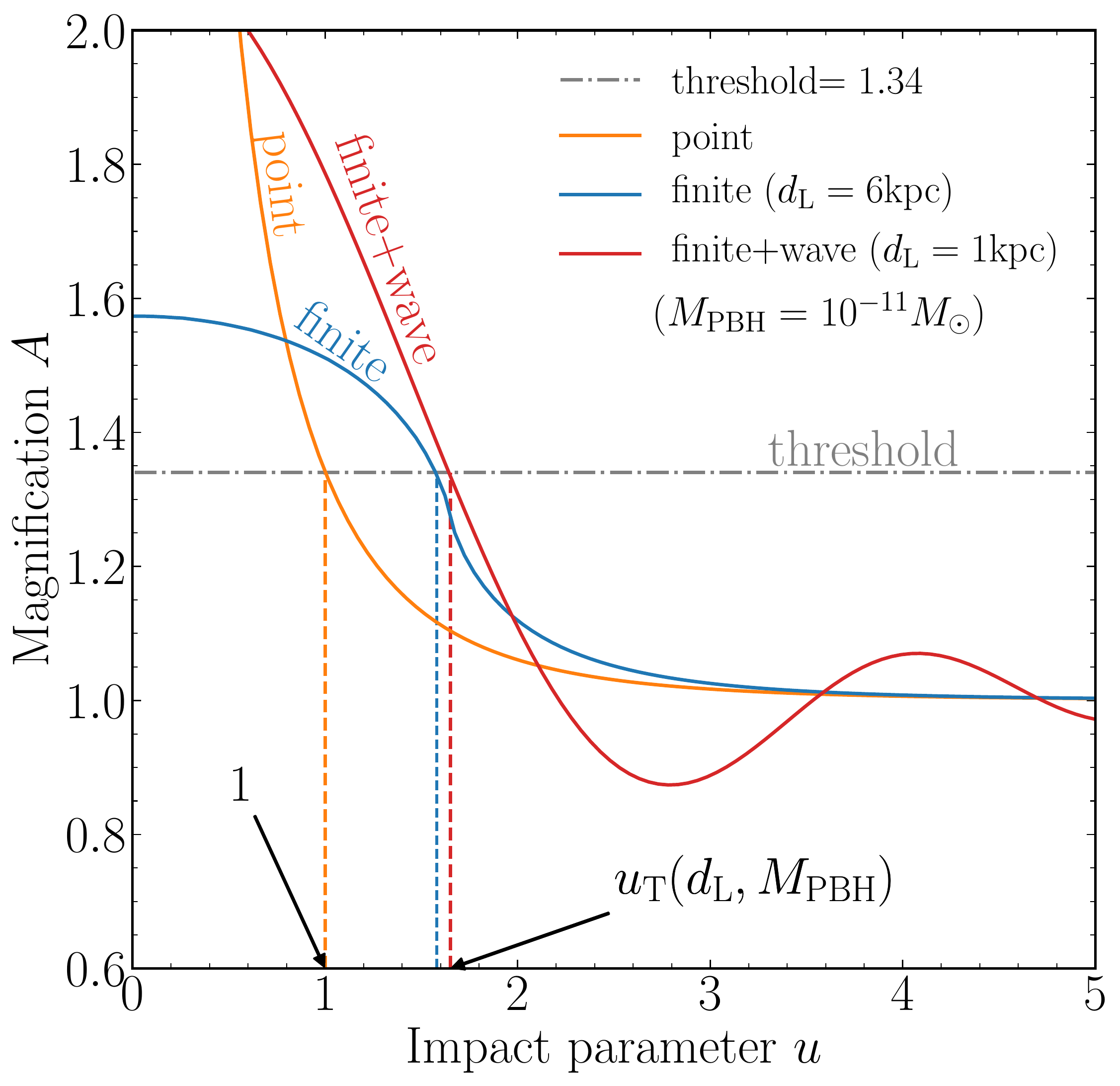}
    \caption{Comparison of the microlensing light curves for a point source or when taking into account the
    finite source size effect and/or the wave optics effect. Here we consider $M_{\rm PBH}=10^{-11}M_\odot$ for the PBH mass.
    The vertical line denotes $A=1.34$, which corresponds to a nominal threshold of lensing magnification leading to a detection of the event in an observation. For this, the vertical lines gives the impact parameter threshold, $u_{\rm T}$, corresponding to the magnification threshold for each case. The lensing magnification $A\ge 1.34$ at $u\le u_{\rm T}$.
}
    \label{fig:threshold}
\end{figure}

\begin{figure}
	\includegraphics[width=1.0\columnwidth]{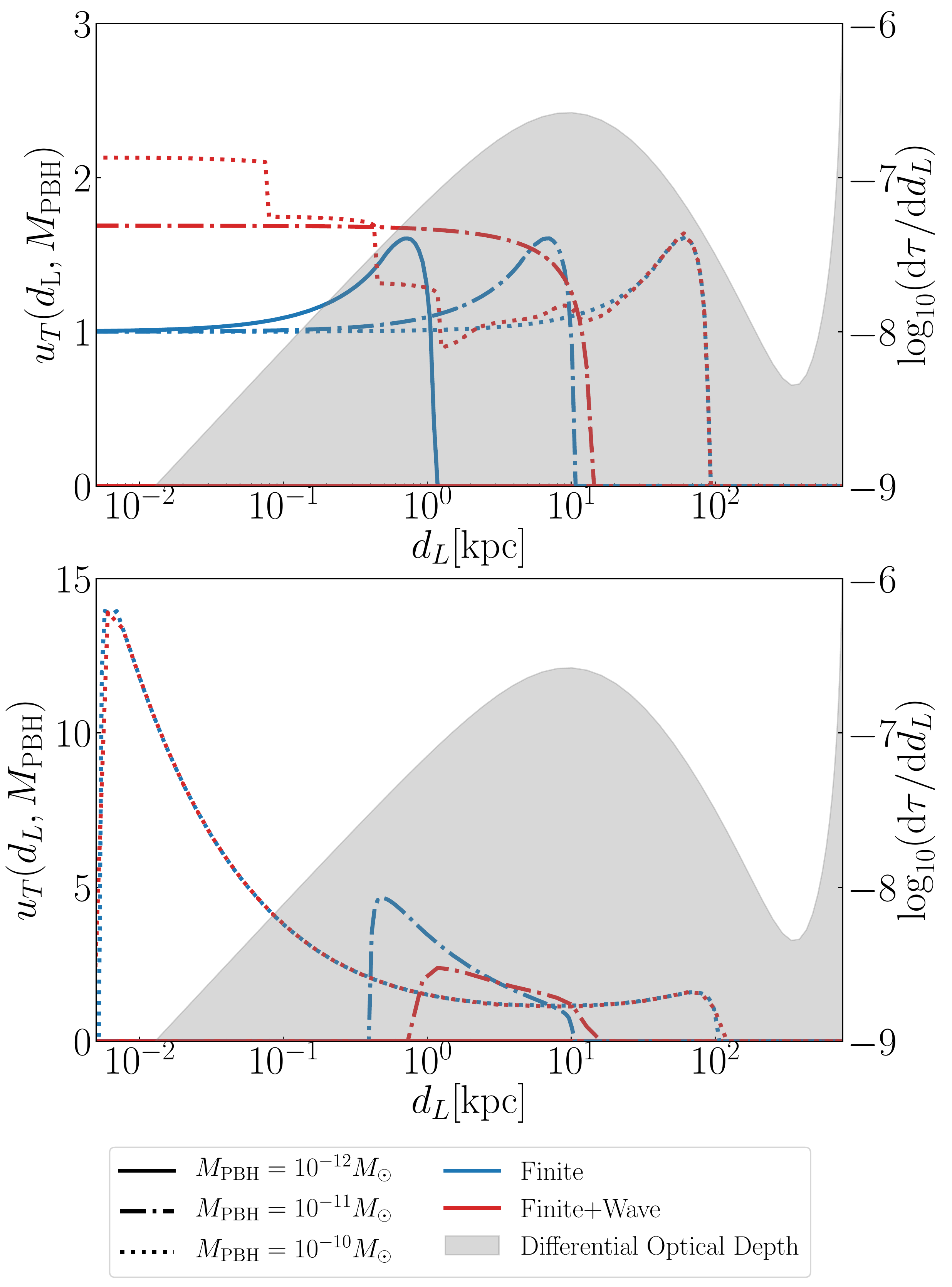}
    \caption{The impact parameter threshold $u_{\rm T}$ as a function of PBH distance, where the lensing magnification for a source star in M31 becomes greater than the detection threshold, $A>1.34$ when $u<u_{\rm T}$ as discussed in the previous figure. Note $u_{\rm T}=1$ for a point source independently of the lens distance. The different curves, as indicated by the legend at bottom, show the results when taking into account the finite source size effect and/or the wave optics effect, for PBHs of different mass scales as indicated. When PBHs are too light, no PBH can cause a detectable microlensing due to the source size and wave effects. This is the case for $M_{\rm PBH}=10^{-12}M_\odot$ ($u_{\rm T}=0$ for this case).
    The shaded region shows the differential contribution to the microlensing optical depth showing how PBHs at each distance contribute the total optical depth (the integrand of Eq.~\ref{eq:optical depth}). The difference between the upper and lower plots is that, for the lower panel, we include the effect of finite exposure time, 90~sec here; when computing a more realistic light curve of microlensing, we compute the lensing magnification averaged over the exposure time assuming a typical velocity of lensing PBH with respect to a source star at each distance (for which we assume a velocity expected from the velocity dispersion of DM halo at each distance). When PBHs become too light, the light curve has a rapid oscillatory feature as a function of separation between lens PBH and source star as indicated in the previous figure, so it causes a more significant average of the lensing magnification.
    }
    \label{fig:uT_comp}
\end{figure}
In this section we study the impact of wave optics effect and finite source size effect on a search of microlensing for source stars in M31 due to intervening PBHs that exist between the Earth and M31, if PBHs make up DM in the Milky Way and M31 halo regions, by a certain mass fraction. To do this we employ the parameters in \citet{Niikuraetal_PBH:17} to model the spatial and velocity distributions of DM including PBHs in the Milky Way and M31 regions.

The optical depth of microlensing for a single source star in M31 due to PBHs is given as
\begin{align}
\tau_{\rm PBH}=\frac{\Omega_{\rm PBH}}{\Omega_{\rm DM}}\int_0^{d_{\rm S}}\diff d_\mathrm{L}~\frac{\rho_{\rm DM}(d_{\rm L})}{M_{\rm PBH}}
\pi R_{\rm E}^2,
\label{eq:optical depth}
\end{align}
where a factor $\Omega_{\rm PBH}/\Omega_{\rm DM}$ gives the mass fraction of PBHs to DM that exists along the line-of-sight direction in the Milky Way and M31 regions, and $\rho_{\rm DM}(d_{\rm L})$ is the mass density distribution of dark matter along the line-of-sight direction up to the source star. Here we assume a single source plane; i.e. we assume that all source stars in M31 are at the same distance, which is a good approximation because the spatial extent of stellar distribution in M31 ($\sim 10~{\rm kpc}$) is very small compared to the distance of M31 (770~kpc). To model the spatial distribution of DM, we employ Navarro-Frenk-White (NFW)
 models \citep{NFW97,Klypinetal:02} to reproduce the flat rotation curve for each of the Milky Way or M31, respectively;
we compute the total DM distribution $\rho_{\rm DM}(r)$ by a sum of the contributions of two NFW profiles at the distance of a lens from the Milky Way center and the M31 center, respectively \citep[see][for details]{Niikuraetal_PBH:17}.
Here, we simply included ``microlensing'' events in consideration if lensing PBH and source star are separated by less than the Einstein radius on the sky. Since $R_{\rm E}^2 \propto M_{\rm PBH}$, the above optical depth does not depend on PBH mass. In the following we will more carefully consider how the detection threshold for a microlensing observation could be changed if taking into account the finite source size effect and/or wave optics effect.

The differential event rate of PBH microlensing for a single source star in M31 is defined \citep[see Eq.~18 in][]{Niikuraetal_PBH:17} as
\begin{align}
	\frac{{\rm d}\Gamma_{\rm PBH}}{{\rm d} \hat{t}}=& 2 \frac{\Omega_\mathrm{PBH}}{\Omega_\mathrm{DM}} \int_0^{d_\mathrm{S}}\diff d_\mathrm{L} \int_0^{u_\mathrm{T}(d_\mathrm{L},M_{\rm PBH})}\!\!\mathrm{d}u_{\rm min} \nonumber\\
	&\times\frac{1}{\sqrt{u_\mathrm{T}^2(d_\mathrm{L},M_{\rm PBH})-u_\mathrm{min}^2}} \frac{\rho_\mathrm{DM}(d_\mathrm{L})}{M_\mathrm{PBH} v_\mathrm{c}^2(d_\mathrm{L})} v^4 \exp\left[-\frac{v^2}{v_\mathrm{c}^2(d_\mathrm{L})}\right],
	\label{eq:eventrate}
\end{align}
where $v=2R_{\rm E}\sqrt{u_{\rm T}^2-u_{\rm min}^2}/\hat{t}$. The units of $\mathrm{d}\Gamma_{\rm PBH}/\mathrm{d}\hat{t}$ is $[{\rm events}/{\rm hours}/{\rm hours}]$ giving the event rate for a single star per unit observation time [hours] per unit light curve timescale ($\hat{t}$) [hours].  For the velocity distribution of DM, therefore PBHs, we assume the isotropic, random virial motion according to the NFW halo at a given radius of lensing PBH from the Milky Way center or M31 center.

The integration with respect to the minimum impact parameter $u_{\rm min}$ in
Eq.~(\ref{eq:eventrate}) is in the range $u_{\rm min}=[0,u_{\rm T}]$, where $u_{\rm T}$ is the threshold impact parameter defined as follows.   We observationally detect a microlensing event of PBH if the magnification of a source star
is large enough and then the light curve of a source star is detected under observational conditions. More exactly speaking,
to estimate the detection sensitivity of microlensing events, it requires detailed simulations of microlensing light curve taking into account observation conditions as well as combinations of model parameters, as done in \citet{Niikuraetal_PBH:17}. This is beyond the scope of this paper. Here we simply assume that a microlensing event can be detected if the maximum magnification is greater than a threshold value $A_{\rm T}=1.34$, which corresponds to the magnification at the impact parameter $u=1$ for a point source under geometrical optics approximation (Eq.~\ref{eq:p-nw-magnification}). Even if we consider the wave optics effect and/or the finite source size effect, we can determine the threshold impact parameter, $u_{\rm T}$, corresponding to $A_{\rm T}=1.34$, once the model parameters (distance to PBH, PBH mass, source size, and optical wavelength) are fixed.
If $0\le u_{\rm min}\le u_T$, the microlensing event has a magnification with $A\ge 1.34$.
The integration over $u_{\rm min}$ in Eq.~(\ref{eq:eventrate}) includes microlensing events with $A\ge 1.34$.

Fig.~\ref{fig:threshold} illustrates how the threshold impact parameter $u_{\rm T}$  is changed when taking into account the finite source size effect and the wave optics effect, assuming typical values of the model parameters. For this particular set of the model parameters, the finite source size effect and the wave effect both increase the threshold value of the impact parameter $u_{\rm T}$ compared to the point source and geometrical optics case ($u_{\rm T}=1$). Hence the effects increase a cross section of the microlensing. The oscillatory feature in the light curve appears at $u>u_{\rm T}$ or $A<A_{\rm T}$ for this case. For other parameters,
such oscillatory feature can appear at $u<u_{\rm T}$.
In this way we can estimate the numeric value of $u_{\rm T}$ as a function of the model parameters.

In Fig.~\ref{fig:uT_comp} we study how the threshold impact parameter varies as a function of the distance to PBH for different PBH mass scales as well as the dependence on the finite source size effect and/or the wave optics effect.
For a point source, $u_{\rm T}=1$ independently of the lens distance. First of all, when taking into account the finite source size effect and/or the wave optics effect, only PBHs at particular distances can contribute the microlensing events. If PBH is too light such as $M_{\rm PBH}\simlt 10^{-11}M_\odot$ as we will show below, PBHs can not cause a detectable microlensing event. 
Such an undetectability of lightest PBH leads to a cutoff in the PBH abundance constraint at mass scales below the critical mass
$M_{\rm cut}\simlt 3.3\times 10^{-12}M_\odot$ ($2.5\times 10^{-12}M_\odot$) when we use a microlensing observation of 
$r$-band($g$-band). 
The step-like feature in the result for $M_{\rm PBH}=10^{-10}M_\odot$ arises from the oscillatory feature in the light curve when taking into account the wave effect, if $u_{\rm T}$ is determined according to the method in Fig.~\ref{fig:threshold}. However, as shown in the lower panel,
a finite exposure time has a significant impact on the threshold calculation, where we used Eq.~(\ref{eq:A_texp}) to estimate the effect; the finite exposure time (90~sec) averages out the oscillatory feature in the light curve, leading to no detectable microlensing event, i.e. $u_{\rm T}\rightarrow 0$, if a lensing PBH is closer to an observer\footnote{In this case, the Einstein crossing time (Eq.~\ref{eq:tE}) becomes shorter for a typical velocity of PBH, so the exposure time average smears out the light curve, leading to a smaller net magnification.}.
Even if $u_{\rm T}$ appears to have a greater value for PBHs at $d_{\rm L}\simlt 10^{-1}~{\rm kpc}$ for $M_{\rm PBH}=10^{-10}M_\odot$, the contribution of PBHs at such small distances to the microlensing event rate is very small as indicated by the gray-shaded region, and therefore the contribution to the event rate is negligible.

\begin{figure}
    \includegraphics[width=1.0\columnwidth]{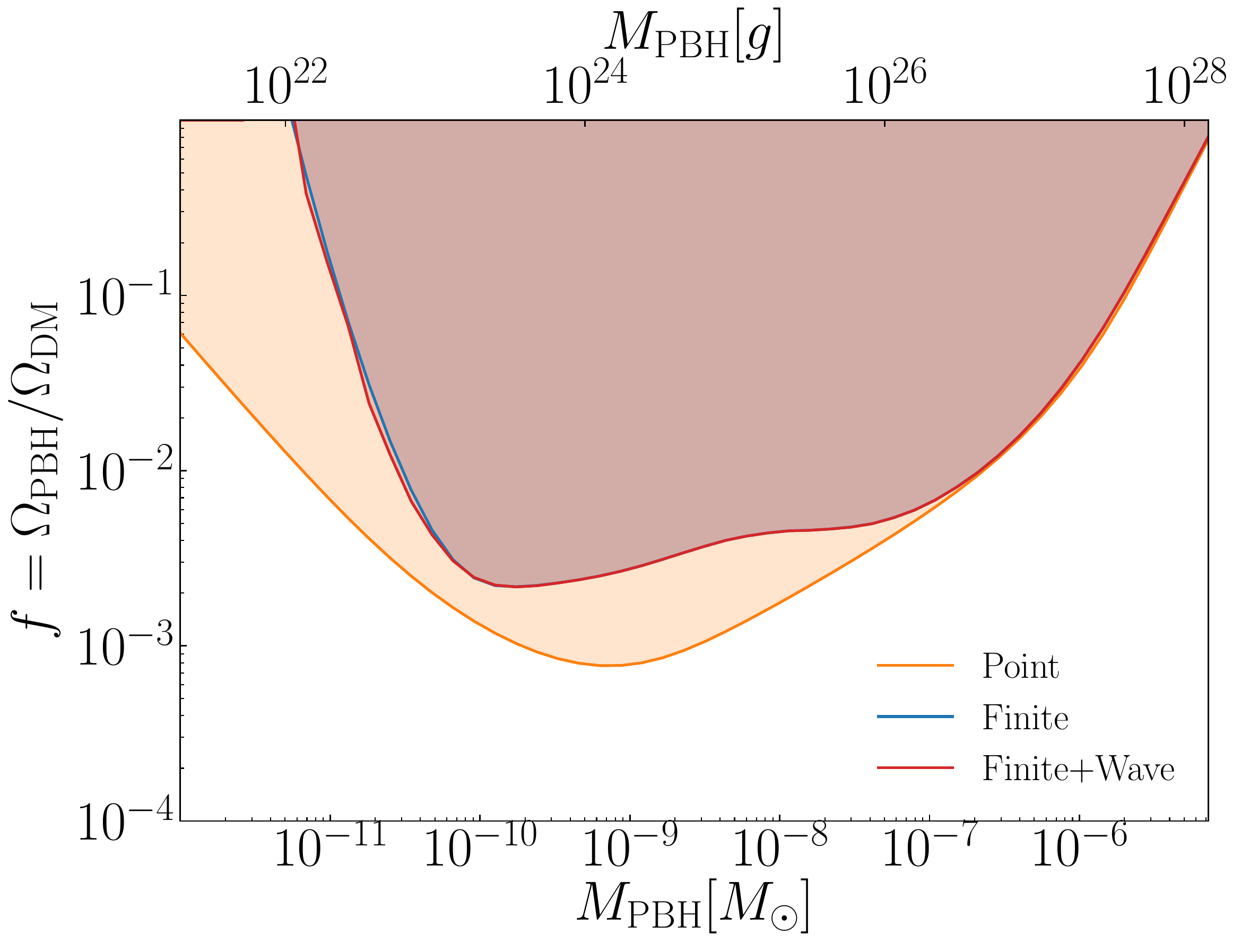}
   \caption{The impact of the finite source size effect and the wave effect on the 95\% C.L. upper bound on the PBH mass fraction to DM in the halo regions of the Milky Way and M31 in \citet{Niikuraetal_PBH:17}, which are based on the $r$-band filter data of M31 taken with the Subaru HSC. Note that the $r$-band filter has 6210\AA~for the central wavelength, and they used $8.7\times 10^7$ stars for the constraints.
   Compared to the point source result, the source size and wave effects cause a sharp cut at $M_{\rm PBH}\simlt 3.3\times 10^{-12}M_\odot$, i.e. no constraint for such light-mass PBHs, as denoted by the vertical line.
   For the finite source size effect we assume the solar radius for all source stars.
   }
   \label{fig:cr_f_r}
\end{figure}
\begin{figure}
	\includegraphics[width=1.0\columnwidth]{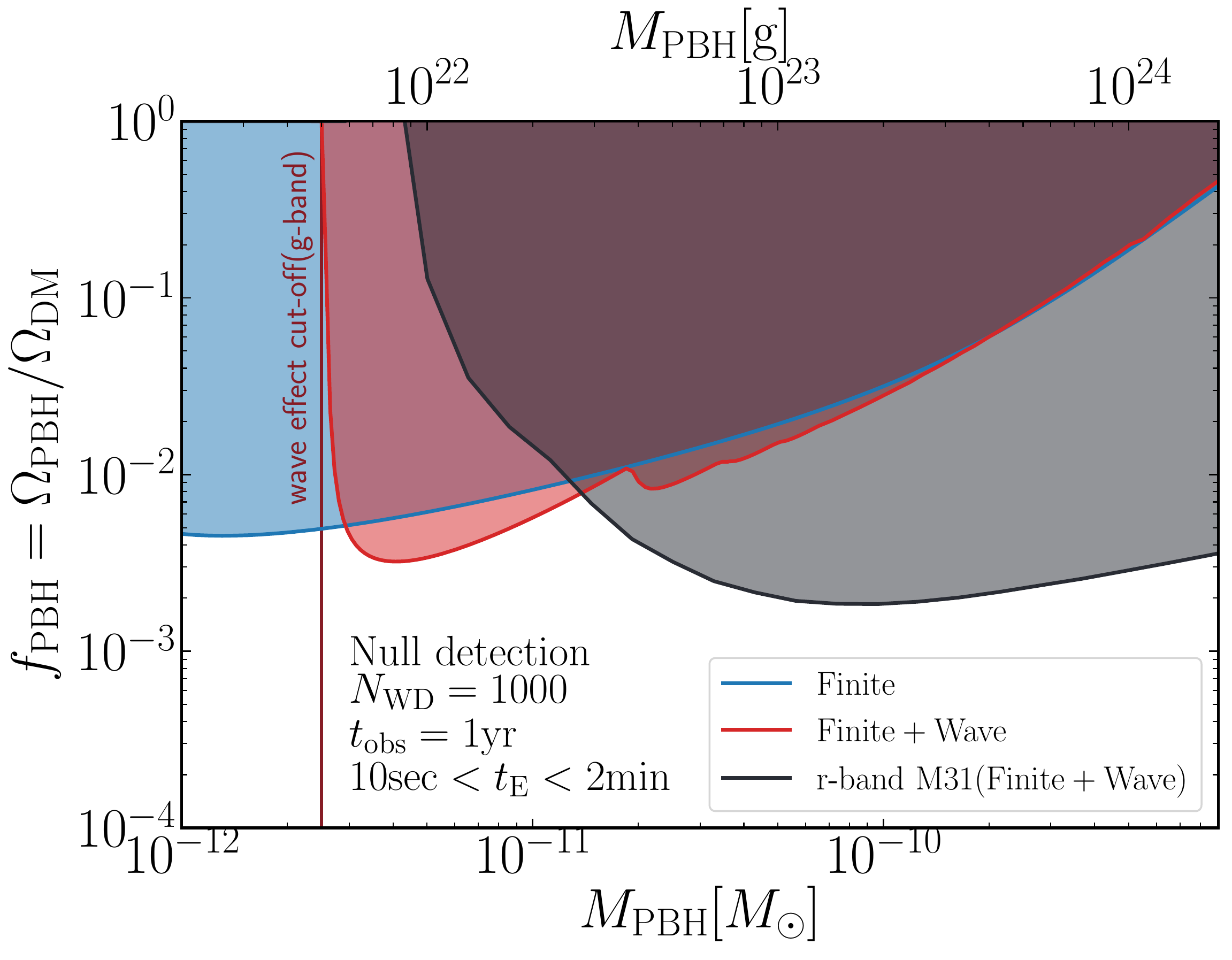}
    \caption{An expected constraint on PBH abundance for a hypothetical dense-cadence monitoring observation of white dwarfs in a $g$-band
    filter (centered at $\lambda=4730$\AA). Here we assume that we can have a monitoring observation for 1000 white dwarfs in LMC,
    that all white dwarfs have the same radius of $0.01R_\odot$,
    that we can perfectly recover the microlensing light curve in the timescale range of $10~{\rm sec}\le t_{\rm E}\le 2~{\rm min}$, if occurs, and $t_{\rm obs}=1~{\rm year}$ for the net, total observation time (see text for details). The red curve shows the expected 95\% C.L. upper limit on the PBH abundance, if no lensing event is detected (i.e. null detection), when taking into account both the wave optics and finite source size effects. The blue curve shows the result when including the finite source size effect alone. Hence the cutoff mass scale at $M_{\rm PBH}\simeq 2.5\times 10^{-12}M_\odot$, as denoted by the vertical line, is due to the wave optics effect for the $g$-band observation. For comparison, the gray curve shows the red curve in Fig.~\ref{fig:cr_f_r}.}
    \label{fig:cr_eff_g}
\end{figure}
Fig.~\ref{fig:cr_f_r} is the main result of this paper. Here we use the microlensing search results from one-night Subaru HSC $r$-filter data of M31 in \citet{Niikuraetal_PBH:17}, where they found only one possible candidate of PBH microlensing compared to many expected events if all DM is made up of PBHs. The results were translated into an upper bound on the abundance of PBHs at each mass scale assuming a monochromatic mass spectrum for PBHs. 
Fig.~\ref{fig:cr_f_r} compares the results for three cases: i) if we assume a point source and ignore the wave optics effect (most optimistic case), ii) if we take into account the finite source size effect, and iii) if we take into account both the finite source size effect and the wave optics effect. Here we assume the solar radius for source size for all stars in M31, and 
include the ``selection'' function of microlensing events that give a probability to recover microlensing events of a given light curve timescale under the Subaru HSC observation conditions (taken from Figure~18 in Niikura~et~al.). The figure shows almost no difference between the results of cases ii) and iii), meaning that the finite source size effect is a dominant effect for low-mass PBHs in $M_{\rm PBH}=[10^{-11},10^{-7}]M_\odot$. In other words, the wave effect is very difficult to distinguish in this kind of statistical study of microlesning search based on the $r$-band data.

\section{Discussion}
\label{sec:discussion}

Fig.~\ref{fig:cr_f_r} clearly shows a fundamental limitation of the PBH constraint from the Subaru $r$-band data; no constraint on PBHs at $M\simlt 10^{-11}M_\odot$.
How can we improve the microlensing constraints on PBH abundance from an optical observation? One possibility is to use a shorter-wavelength filter for the microlensing search because the ratio of PBH's Schwarzschild radius to the light wavelength becomes greater, which reduces the impact of wave optics effect. For example, if we can use a $g$-band filter whose central wavelength
($\lambda=4730$\AA) is shorter than that of $r$-band filter ($\lambda=6210$\AA) we have considered, we could still monitor many stars. In particular, white dwarfs are bluer than main sequence stars, and have a smaller radius which is typically smaller than 
the solar radius by a factor of 100. This reduces the impact of finite source size effect. However, the abundance of white dwarfs is smaller than that of main sequence stars, by a factor of 10 as implied from the MOA or OGLE microlensing events for stars in the Large Magellanic Cloud (LMC) or the Galactic bulge region \citep{Sumietal:03,2017Natur.548..183M,2019PhRvD..99h3503N}. In addition, white dwarfs are much fainter than main sequence stars (by more than 5 magnitudes than main sequence stars\footnote{E.g., see 
\url{ http://sci.esa.int/gaia/60209-white-dwarfs-in-gaia-s-hertzsprung-russell-diagram/}}). So a large-aperture telescope is needed, or we need to monitor white dwarfs at closer distances than in M31, e.g. white dwarfs in the Galactic bulge or LMC so that white dwarfs are bright enough to be detected by a reasonably large-aperture telescope such as the 8.2m Subaru telescope.

To explore microlensing event(s) due to
PBHs in the mass range
$M_{\rm PBH}\simlt 10^{-11}M_\odot$, we need a much denser-cadence observation of white dwarfs than done in the Subaru $r$-band 2~min-cadence observation of \citet{Niikuraetal_PBH:17} (90~sec exposure plus 30~sec exposure), in order to well sample a microlensing light curve of much shorter timescale than 2~min. The upcoming Large Synoptic Survey Telescope 
(LSST)\footnote{\url{https://www.lsst.org}}
will allow for 2~sec readout time that is much shorter than 30~sec of the Subaru HSC. Furthermore, 
if a large-format CMOS image sensor is available for a large aperture telescope, it will allow a much faster readout imaging of source stars including white dwarfs. Here we assume that we can have a monitoring observation for a sample of 1000 white dwarfs, compared to $8.7\times 10^7$ stars in M31 in \citet{Niikuraetal_PBH:17}, using a sufficiently dense cadence observation with a 
$g$-band filter at a large aperture telescope. We assume that the 1000 white dwarfs are at 50~kpc in distance (e.g., LMC)
so that those are bright enough, and that all the white dwarfs have the same radius,  $R_{\rm S}=0.01R_\odot$ for simplicity.
Then we assume that we can well sample a light curve of microlensing, if occurs, in the range of $10~{\rm sec}\le t_{\rm E}\le 2~{\rm min}$ for the typical light curve timescale. More exactly we assume that we can perfectly recover a microlensing event of the timescale range, if occurs. Finally we assume we have a 1-year amount of data for the net observation time. Such a telescope/detector would be interesting to explore, and seems feasible within next 10 years or so.  Here we simply use the same models in \citet{Niikuraetal_PBH:17} to study forecasts for such a microlensing search (so we assume the white dwarf sample is in the direction to M31).
In addition, a microlensing of white dwarfs would be more suitable to search for the wave optics effect due to the relatively weaker impact of the finite source size effect.

Fig.~\ref{fig:cr_eff_g} shows the expected upper bound on the PBH abundance from the $g$-band microlensing search we described above, assuming no secure microlensing event (i.e. null detection). Compared to Fig.~\ref{fig:cr_f_r}, such a $g$-band dense-cadence monitoring observation can constrain PBHs at lighter mass scales, down to a few times $10^{-12}M_\odot$. The cut off at the low PBH mass is from the wave optics effect; PBHs in the lighter mass scales cannot cause microlensing.
This result can be understood as follows. The expected number of microlensing events is roughly given as $N_{\rm exp}\sim N_{\rm S}\times \tau\times (t_{\rm obs}/t_{\rm E})$, where
$N_{\rm S}$ is the number of source stars (here white dwarfs), $\tau$ is the optical depth of PBH microlensing for a single source star, and $t_{\rm obs}$ is the net observation time. The optical depth $\tau\simeq 10^{-6}$ for the Milky Way halo if PBHs make up all DM in the Milky Way halo region. Hence, if we seach for a microlensing event of $10~{\rm sec}$ timescale, the expected number is
$N_{\rm exp}\sim 10^3\times 10^{-6}\times (3\times 10^7~{\rm sec})\times (10~{\rm sec})^{-1}\sim 3000$. If we cannot find any event, i.e. null detection, it gives an upper limit that such PBHs are not allowed by more than $1/3000=0.003$ for the mass fraction of PBHs to DM, which roughly reproduces the result in Fig.~\ref{fig:cr_eff_g}. More quantitatively, in the figure we took into account the distribution of light curve timescales due to the velocity distribution of PBHs in the Milky Way halo region.

\section{Conclusion}
\label{sec:conclusion}

In this paper, we studied the effects of wave optics and finite source size on the optical microlensing search for stars in M31 due to PBHs that would exist in the Milky Way and M31 halos if DM is made up of PBHs. If PBH is in mass scales of $M_{\rm PBH}\simlt 10^{-10}M_\odot$, its Schwarzschild radius ($r_{\rm Sch}$) becomes comparable with or shorter than the optical filter wavelength (e.g. the $r$-band filter, centered at $6210$\AA), and the wave effect on microlensing needs to be considered for such PBHs. For PBHs with
$r_{\rm Sch}\ll \lambda$, even PBHs cannot bend the path of optical light, so causes no microlensing magnification, which gives the fundamental limit for an optical microlensing search ($M_{\rm PBH}\simlt 10^{-11}M_\odot$ for the optical $r$-band filter). Nevertheless,
if we can find a secure microlensing event for PBH with $\sim 10^{-10}M_\odot$ in an optical filter observation, it is a smoking-gun evidence of PBHs because any other astrophysical object cannot have such a tiny Schwarzschild radius (their physical size is much larger than the optical wavelength). This is an interesting possibility to explore from an actual observation. However, we showed that the finite source size effect is equally important or even more significant, if the microlensing observation targets main-sequence stars, and likely erases characteristics signatures of the wave effect in the microlensing light curve. As the main result of this paper, we studied the impact of wave effect and finite source size effect on the PBH constraints obtained from the Subaru HSC microlensing search for stars in M31 in \citet{Niikuraetal_PBH:17}. We showed that the effects are significant, and the finite source size effect is a dominant effect compared to the wave effect, if source stars have a size of the solar radius as expected for main sequence stars
(see Fig.~\ref{fig:cr_f_r}). Nevertheless, upcoming wide-area surveys such as LSST, Euclid and WFIRST would be very powerful to search for microlensing events due to PBHs over a wide range of mass scales, so the results shown in this paper are relevant for microlensing search from the upcoming surveys.

In order to ``detect'' the wave optics effect of microlensing in an optical observation, we need to use a source star which has a smaller size. Such a source is white dwarf, which has a smaller size (typical a few \% the solar radius), and can be observed in optical wavelengths. However, the number of white dwarfs are smaller than that of main sequence stars, by a factor of 10, and white dwarfs are much fainter than main sequence stars, by more than 5 magnitudes (a factor of 100 in the flux). Hence the microlensing search for white dwarf sources is challenging, but it is not impossible. For example, a CMOS sensor detector would be promising, compared to CCD camera, because it allows a much denser-cadence monitoring observation of source stars and therefore enables to search for ultrashort timescale microlensing events (the readout time of Subaru HSC is 30~sec). We discussed that such a monitoring observation in $g$-band filter can improve the PBH abundance in the mass scales, $M\simlt 10^{-11}M_\odot$, and therefore such a CMOS camera with large field-of-view at a large-aperture telescope would be powerful to further explore PBH signatures.

\section*{Acknowledgements}
We thank Nick Kaiser and Misao Sasaki for pointing out the importance of wave optics effect for the HSC M31 microlensing search during the seminar MT gave at the Yukawa Institute for Theoretical Physics, Kyoto University. 
This work was initiated by the suggestion. We also thank
Hiroko Niikura, Chris Hirata and Ryuichi Takahashi for useful discussions.
This work was supported in part by World Premier International
Research Center Initiative (WPI Initiative), MEXT, Japan, and JSPS
KAKENHI Grant Number JP15H03654,
JP15H05887, JP15H05893, JP15K21733, and 19H00677.

\bibliographystyle{mnras}
\bibliography{refs}

\label{lastpage}
\end{document}